\documentclass[prl,aps,amssymb,showpacs,twocolumn]{revtex4-1}
\usepackage{amsmath}
\usepackage{amssymb}
\usepackage{amsthm}
\usepackage{amsfonts}
\usepackage{listings}
\usepackage{enumerate}
\usepackage{latexsym}

\usepackage{psfrag}

\usepackage{bm}
\usepackage{graphicx}
\usepackage{subfigure}

\newcommand{\beq}{\begin{equation}}
\newcommand{\eneq}{\end{equation}}

\newcommand{\up}{\uparrow}
\newcommand{\dw}{\downarrow}

\input{epsf}

\begin{document}

\tolerance 10000

\newcommand{\vk}{{\bf k}}


\title{Realizing topological Mott insulators from the RKKY interaction}

\author{Tianhan Liu$^{1,2}$}
\author{Beno\^{\i}t Dou\c{c}ot$^1$}
\author{Karyn Le Hur$^2$}
\affiliation{$^1$ Sorbonne Universités, Universit\'e Pierre et Marie Curie, CNRS, LPTHE, UMR 7589 , 4 place Jussieu, 75252 Paris Cedex 05\\ $^2$ Centre de Physique Th\'eorique, \'Ecole polytechnique, CNRS, Universit\'e Paris-Saclay, F-91128 Palaiseau, France}

\begin{abstract}
We engineer topological insulating phases in a fermion-fermion mixture on the honeycomb lattice, without resorting to artificial gauge fields or spin-orbit couplings and considering only local interactions. Essentially, upon integrating out the fast component (characterized by a larger hopping amplitude) in a finite region of dopings, we obtain an effective interaction between the slow fermions at half-filling, which acquires a Haldane mass with opposite parity in the two valleys of the Dirac cones, thus triggering a quantum anomalous Hall effect. We carefully analyze the competition between the induced Semenoff-type mass (producing charge density wave orders in real space) versus the Haldane mass (quantum anomalous Hall phase), as a function of the chemical potential of the fast fermions. If the second species involves spin-1/2 particles, this interaction may induce a quantum spin Hall phase. Such fermion-fermion mixtures can be realized in optical lattices or in graphene heterostructures.
\end{abstract}

\date{\today}

\maketitle

The quest for topological phases in the absence of a net uniform magnetic field, has attracted a great amount of attention recently in the field of condensed matter physics, in connection with spin-orbit coupling and artificial gauge fields \cite{KaneHasan,QiZhang,Dalibard,Goldman}. The realization of such phases has become important due to their physical properties such as edge transport and potential applications for spintronics \cite{Pesin}. The HgTe quantum well and three-dimensional bismuth analogs have been a perfect area for the quantum spin Hall effect and topological band insulators \cite{kane-mele,Bernevig_Zhang,Konig,Hasan}. In addition, the quantum anomalous Hall effect and its version on the honeycomb lattice, the Haldane model \cite{Haldane}, have been observed with photons \cite{RaghuHaldane,Soljacic}, cold atom systems \cite{Esslinger} and magnetic topological insulators \cite{mti}. Synthetic gauge fields and spin-orbit couplings are currently vastly investigated in optical lattices \cite{atoms} and photon analogs \cite{photons}. Engineering topological phases through interactions is also interesting on its own. Interactions may also localize the charge, through Mott physics. The transition between topological band insulator and Mott phase has been largely addressed  \cite{pesin-10np376,RachelLeHur, Assad, Greg,Walter,Tianhan}. An example of topological band insulators induced by interactions, resulting in topological Mott insulators, has also been proposed by Raghu {\it et al}. \cite{Raghu} on the honeycomb lattice. This scenario requires however that the next-nearest-neighbour interaction exceeds the nearest-neighbour repulsion \cite{Raghu,Maria,Marcel}. In addition, recent numerical works have questioned the existence of a topological phase within this model \cite{Herbut,t_V1_V2}. In this paper, we envision a fermion-fermion mixture comprising local interactions, for the realization of such topological Mott insulators. The idea here is that one copy of the mixture is much faster than the other, inducing an exotic Ruderman-Kittel-Kasuya-Yosida (RKKY) interaction \cite{RKKY} on the partner copy. We will show below that the long-range aspect of the RKKY interaction can contribute to frustration of the charge density wave orders.

\begin{figure}[h]
\includegraphics[width=0.75\linewidth]{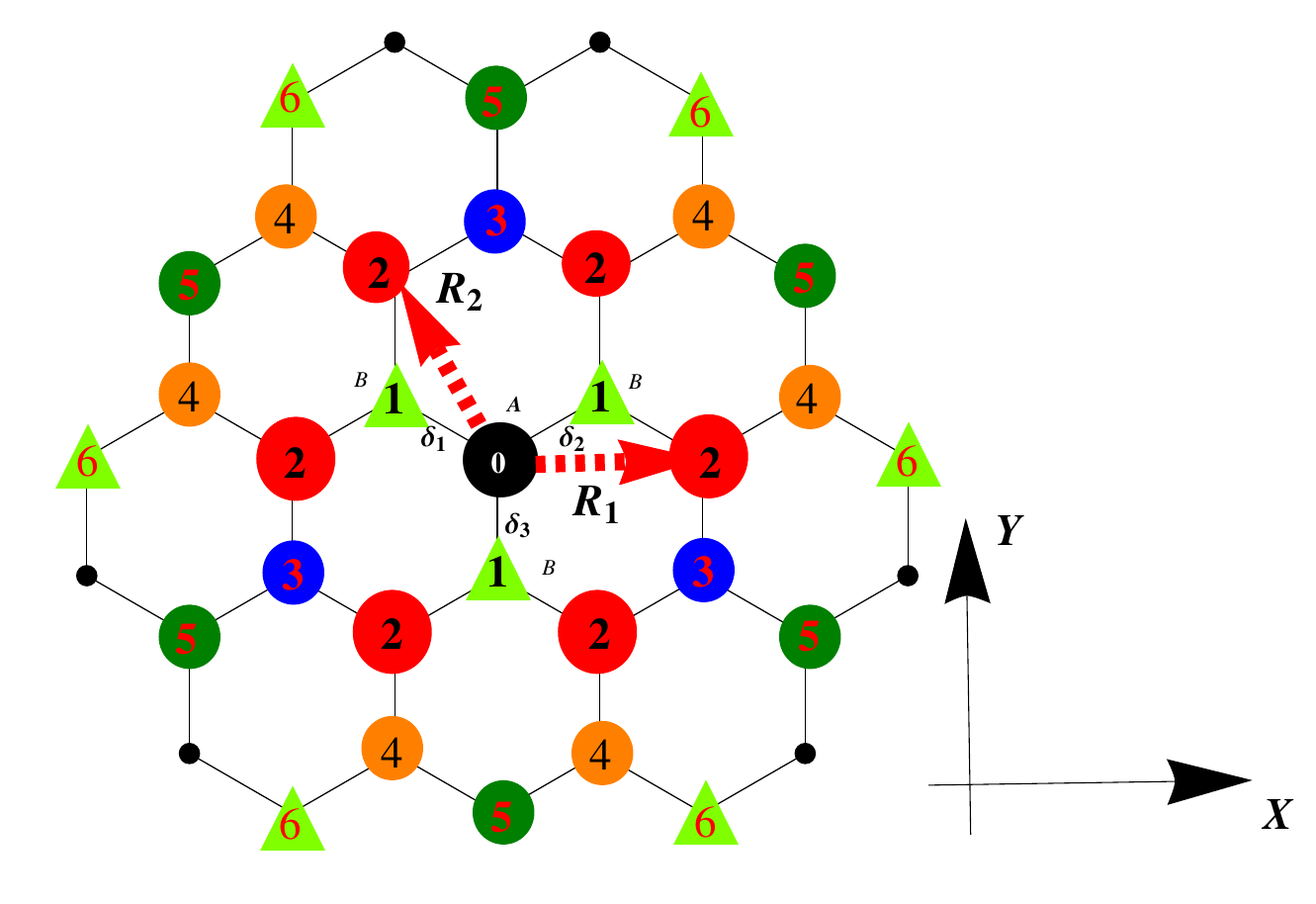}
\includegraphics[width=0.7\linewidth]{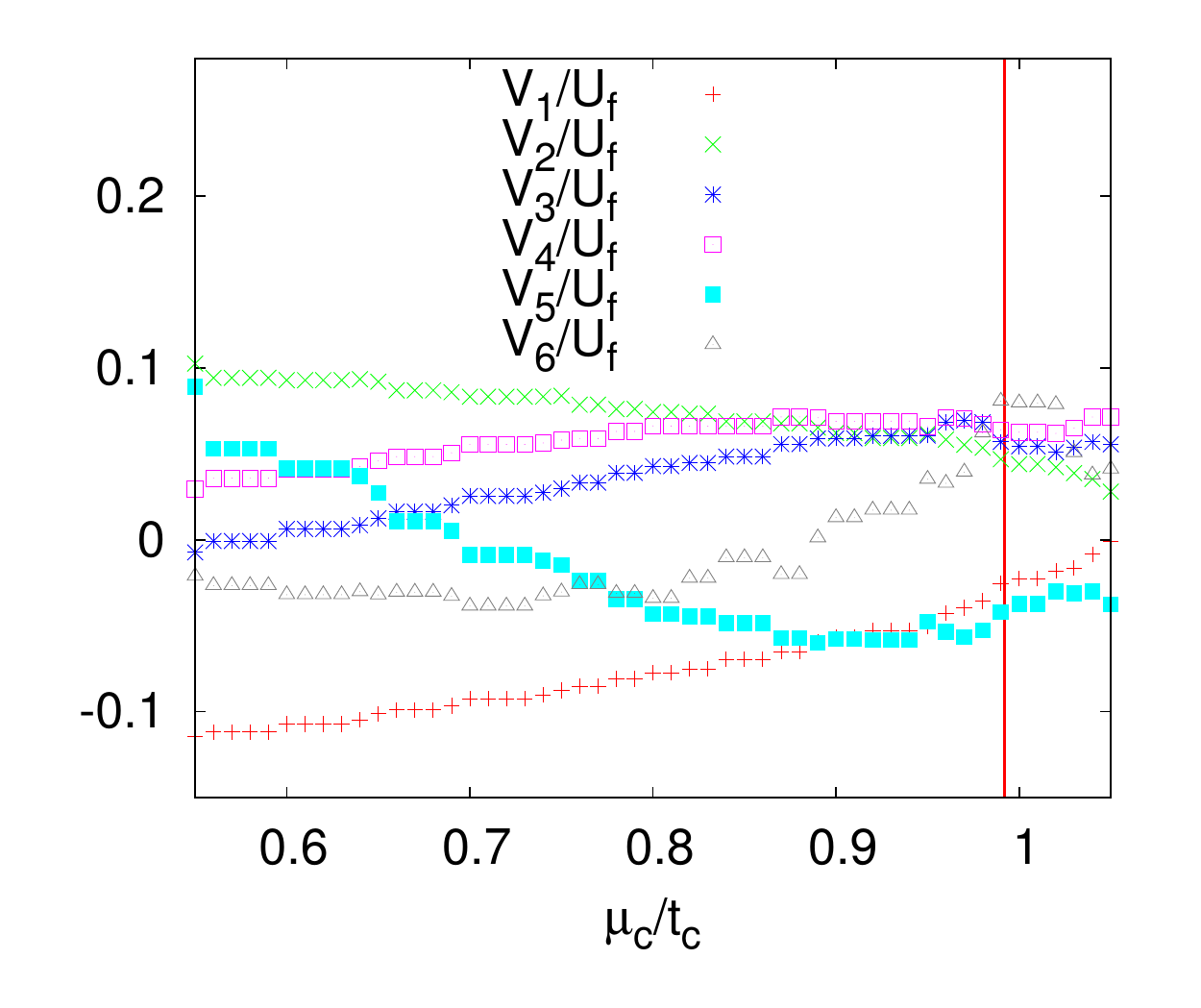}
\caption{(Color online) Upper panel: Graphene honeycomb lattice with the two sublattices $A$ and $B$. We introduce $1, 2, 3... 6$ to label the six nearest-neighbour partners of site 0. Lower panel:The ratio of interaction $V_i/U_f$ ($i=1,2..6$) as a function of the chemical potential of the fast fermions, where $i$ refers to the definition of sites in the upper panel with $U_f$ the on-site interaction.}\label{rkky_08}
\end{figure}

This class of topological Mott insulators, which refers to topological band insulators induced by interaction effects \cite{Raghu}, must be distinguished from the other class of (three-dimensional) topological Mott insulators, which corresponds to a Mott phase where the spin sector is embodied by a band structure of a topological band insulator giving rise to purely neutral spin edge modes \cite{pesin-10np376}. Known examples of topological Mott phases also exist in one dimension \cite{Haldanespin,Affleck}. Topological aspects of Kondo materials have been recently addressed theoretically and observed experimentally \cite{Dzero,Mesot,Fisk}.

Cold atom graphene systems \cite{Leticia} and heterostructures on the honeycomb lattice may be relevant for experimental realizations and the adjustability of the interactions among the mixture \cite{Ketterle,Hansch} will allow for the exploration of various quantum phases. Mixtures of ultra-cold atoms have already attracted some attention for the realization of topological phases and supersolid phases \cite{mixture}. The RKKY interaction has already been shown to be efficient to engineer topological superconductors \cite{TSC}.

\emph{Model and Notations}- We propose a model on the honeycomb lattice with two species of fermions ($c$ and $f$) among which there is one {\it spinless} species of fermions $c$ and one spin-1/2 (or spinless) species $f$. The two species of fermions are coupled together via a repulsive interaction characterized by the coupling constant $g_{cf}$. In the case of spin-1/2 $f$ fermions, we also consider the effect of an on-site Hubbard interaction $U_f$, resulting in the Hamiltonian:
\begin{eqnarray}
\begin{split}
H=&-t_f\sum_{\left<i,j\right>, \sigma}f_{i\sigma}^{\dagger}f_{j\sigma}+\mu_f\sum_{j,\sigma}f_{j\sigma}^{\dagger}f_{j\sigma}+U_{f}\sum_if_{i\up}^{\dagger}f_{i\up}f_{i\dw}^{\dagger}f_{i\dw}\\
&-t_c\sum_{\left<i,j\right>}c_i^{\dagger}c_j+\mu_c\sum_jc_j^{\dagger}c_j+\sum_{j, \sigma}g_{cf}f_{j\sigma}^{\dagger}f_{j\sigma}c_j^{\dagger}c_j .
\end{split}
\end{eqnarray}

We take the lattice spacing to be $1$. The three vectors indicating the nearest neighbors are $\pmb{\delta}_1=(-\frac{\sqrt{3}}{2},  \frac{1}{2})$, $\pmb{\delta}_2=(\frac{\sqrt{3}}{2},  \frac{1}{2})$, and $\pmb{\delta}_3=(0,  -1)$. Both the fermions $f$ and fermions $c$ have the band structure of a graphene system. We can diagonalize the band structure of fermions $c$ on the honeycomb lattice by introducing the isospin of the two sublattice fermions $\Phi_c=(c_{A{\bf k}},c_{B{\bf k}})^T$:
\begin{eqnarray}
\begin{split}
H_c&=-t_c\sum_{\left<i,j\right>}c_i^{\dagger}c_j+\mu_c\sum_ic_i^{\dagger}c_i=\sum_k\Phi_c^{\dagger}\mathcal{H}_k^c\Phi_c\\
\mathcal{H}_k^c&=\left(\begin{array}{cc} \mu_c & -t_c g^*({\bf k}) \\ -t_c g({\bf k}) & \mu_c \end{array}\right),
\end{split}
\end{eqnarray}
in which $g({\bf k})=\sum_{j=1, 2, 3}e^{i{\bf k}\cdot {\pmb{\delta}_j}}$.  We obtain two bands of fermions with the energy levels: $\epsilon_{\pm}({\bf k})=\mu_c\pm t_c|g({\bf k})|$ and the related annihilation operators $\Phi_{c\pm}=\frac{1}{\sqrt{2}}(c_{A{\bf k}}\pm c_{B{\bf k}}e^{i\phi_{\bf k}})$, $\phi_{{\bf k}}=\mathrm{arg}[g({\bf k})]$.

Below, we consider the case where the fermions $c$ are much faster than the fermions $f$ ($t_c\gg t_f$), justifying the formal Gaussian integration of the fermions-$c$. For simplicity, we assume that the fermions $c$ are spin-polarized (one-component).  The induced dynamical RKKY susceptibility involves the Lindhard function:
\begin{equation}\label{susc_rkky}
\chi_{IJ}(\Omega, \mathbf{q}, \mu_c)=-\sum_{\mathbf{k}}\lim_{\eta\rightarrow0}\frac{n_f[\epsilon_-(\mathbf{k}+\mathbf{q})]-n_f{\epsilon_-(\mathbf{k})}}{\Omega+\epsilon_-(\mathbf{k}+\mathbf{q})-\epsilon_-(\mathbf{k})+i\eta}\alpha_{IJ}(\mathbf{k},\mathbf{q}),
\end{equation}
in which $n_f[\epsilon_-({\bf k})]$ is the Fermi-Dirac distribution of the fermions $c$ with energy of the lower band and $IJ$ are the indices for the sublattices $A$ and $B$. $\alpha_{AB}(\mathbf{k}, \mathbf{q})=e^{i(\theta_{\mathbf{k}+\mathbf{q}}-\theta_{\mathbf{k}})}$, $\alpha_{BA}(\mathbf{k}, \mathbf{q})=e^{-i(\theta_{\mathbf{k}+\mathbf{q}}-\theta_{\mathbf{k}})}$ and $\alpha_{AA}=\alpha_{BB}=1$.  We denote the susceptibility on the same lattice as $\chi_{II}(\Omega, \mathbf{q}, \mu_c)$ and the susceptibility on the 
different sub-lattices as $\chi_{AB}(\Omega, \mathbf{q}, \mu_c)$ and $\chi_{BA}(\Omega, \mathbf{q}, \mu_c)^*$.

\begin{figure}[t]
\includegraphics[width=0.495\linewidth]{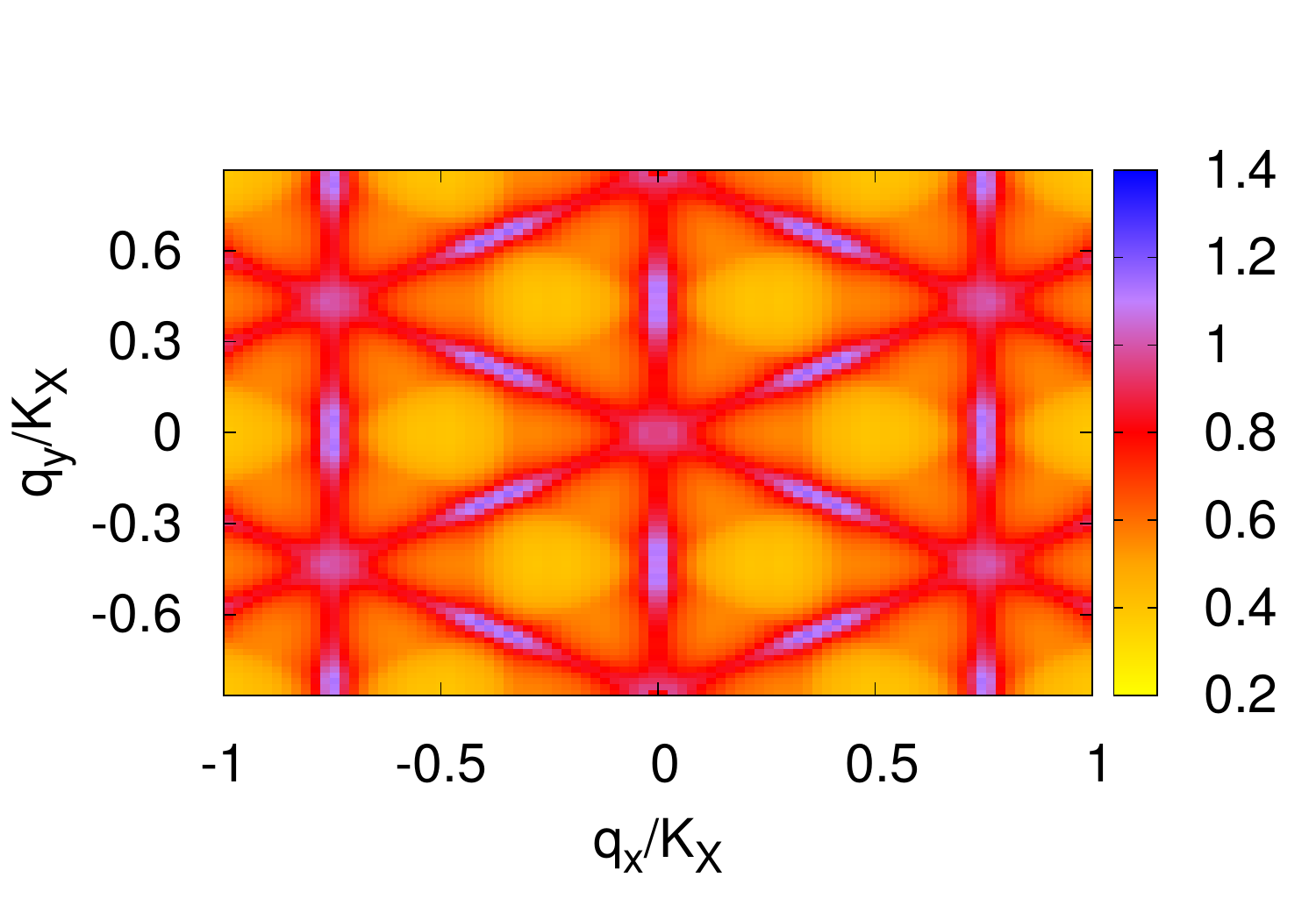}
\includegraphics[width=0.45\linewidth]{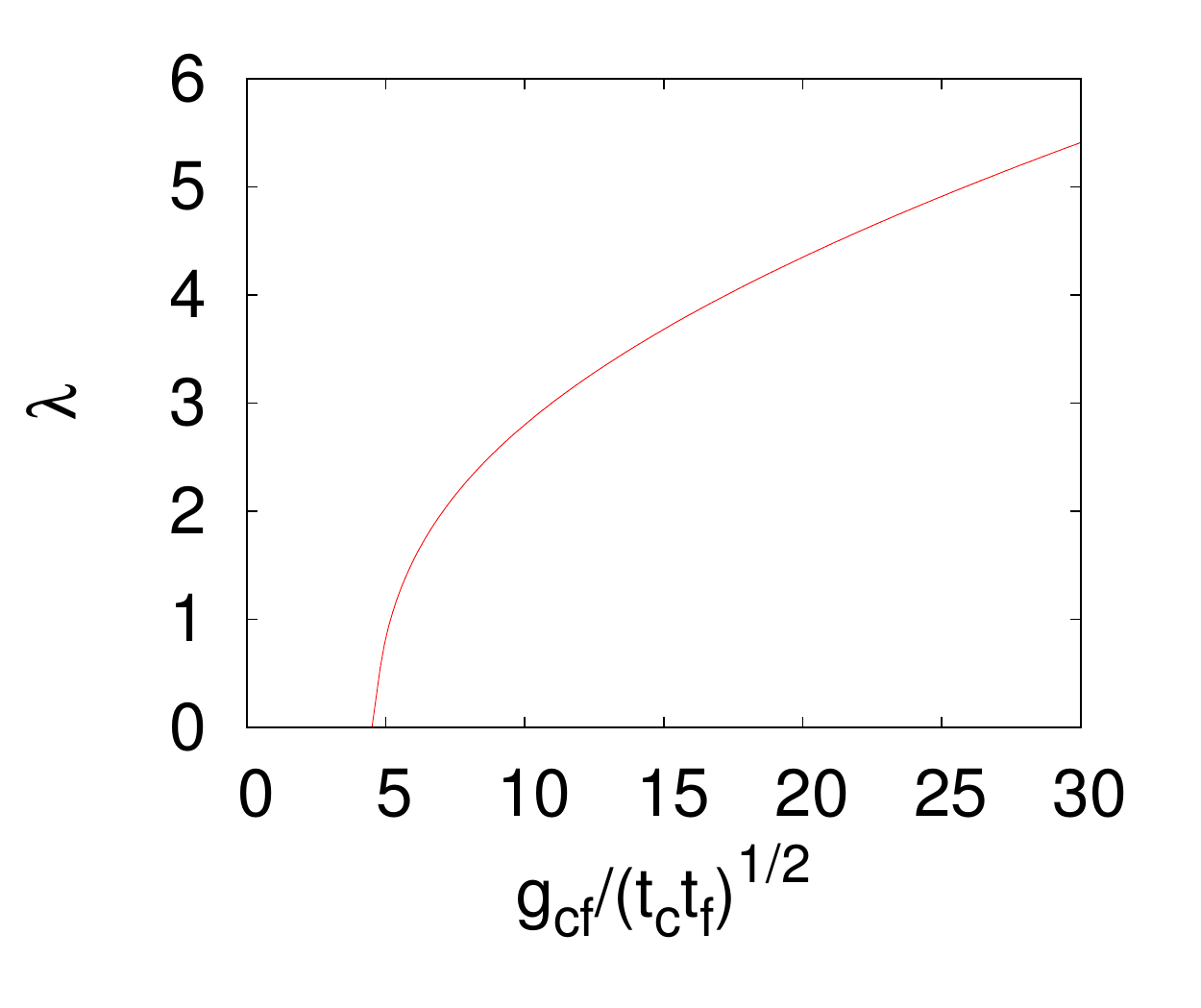}
\caption{(Color online) Left panel: The static RKKY susceptibility on the same sublattice $\chi_{II}(0, \mathbf{q}, \mu_c=0.992t_c)$ as a function of $\mathbf{q}=\frac{\mathbf{Q}}{2}$ as defined in Eq.~(\ref{susc_rkky}).  Right panel: The amplitude of the spontaneous spin-orbit coupling $\lambda$ as a function of $g_{cf}$ when $ \mu_c=0.992t_c$. }\label{rkky_09}
\end{figure}

\emph{Green's function approach}- The RKKY interaction between the fermions $f$, depends strongly on $\mu_c$, which we can adjust. One example is when the static RKKY susceptibility peaks at wave vectors that are nesting vectors in the graphene system close to quarter-filling ($\mu_c=0.992t_c$) with Van Hove singularities (Fig.~\ref{rkky_09}). We have also represented the RKKY interaction in the direct space as shown in Fig.~\ref{rkky_08} upper panel as a function of chemical potential. A similar two-fluid model has been previously proposed on the honeycomb lattice \cite{Refael}, however our model insists on the honeycomb band structure of the fast fermion $c$: this leads to an RKKY interaction with long range interaction and negative nearest-neighbour interaction, which suppresses other orders such as charge density wave and Semenoff mass \cite{Semenoff}. To illustrate these two points, we first proceed to write down the Green function for the slow fermions $f$, and the RKKY interaction induced by the fermion $c$ will imediate an interaction with the fermion $f$. To evaluate the influence of the RKKY interaction on the (bare) Green's function $G_0(\omega, \mathbf{k})$ of a fermion-$f$, we use a standard Hartree-Fock decoupling of the interaction. If we write the spinor $\Psi_{fk}^{\dagger}=(f_{kA}^{\dagger}, f_{kB}^{\dagger})$, then the Green's function and the Schwinger-Dyson equation for the fermion $f$ will be represented as:
\begin{widetext}
\begin{eqnarray}\label{action} 
\begin{split}
G_f(\omega, \mathbf{k})^{-1}_{IJ}&=G_0(\omega, \mathbf{k})^{-1}_{IJ}-i\sum_{\Omega, \mathbf{q}}\frac{g_{cf}^2}{2}\chi_{IJ}(\Omega, \mathbf{q}, \mu_c)G_f(\omega+\Omega, \mathbf{k}+\mathbf{q})_{JI}\\
&\simeq G_0(\omega, \mathbf{k})^{-1}_{IJ}+\sum_{\mathbf{q}}\frac{g_{cf}^2}{2}\chi_{IJ}(0, \mathbf{q}, \mu_c)[2Pf_{JI}(\mathbf{k}+\mathbf{q})-\delta_{JI}]\\
G_0(\omega, \mathbf{k})^{-1}&=\omega-t_f(\tau_x \text{Re}+\tau_y \text{Im})g(\mathbf{k}), 
\end{split}
\end{eqnarray}
\end{widetext}
in which $\tau_x$, $\tau_y$ and $\tau_z$ are Pauli matrices in sub-lattices subspace.  We have checked that the peaks of $\chi_{IJ}(\Omega, \mathbf{q}, \mu_c)$ vary less than $20\%$ when $|\Omega|<0.3t_c$, we have therefore made the adiabatic approximation in the second equality of Eq.~(4), which consists of replacing the dynamical RKKY susceptibility by the static susceptibility: $\chi_{IJ}(\Omega, \mathbf{q}, \mu_c)\simeq \chi_{IJ}(0, \mathbf{q}, \mu_c)$. $Pf_{JI}(\mathbf{k}+\mathbf{q})$ is the projector to the lower band for the fermion $f$. 

An important point is that the Hartree term and certain contributions from the Fock term will change the chemical potential of the fermion $f$. Here, we evaluate:
\begin{eqnarray}
\begin{split}
\tilde{\mu}_f=&\mu_f+g_{cf}^2[\text{Re}\chi_{II}(0, 0, \mu_c)-\sum_{q\neq0}\text{Re}\chi_{II}(0, q, \mu_c)]\left<n_k\right>\\
&+\frac{g_{cf}^2}{2}\sum_{q}\text{Re}\chi_{II}(0, q,  \mu_c).
\end{split}
\end{eqnarray}

We find numerically that $\tilde{\mu}_f-\mu_f$  is negligible (ranging from  $0.03t_f$ to $0.09t_f$) for $g_{cf}<20\sqrt{t_ct_f}$. This justifies our consideration of the half-filled case for the fermion $f$.

\begin{figure}[th]
\includegraphics[width=0.49\linewidth]{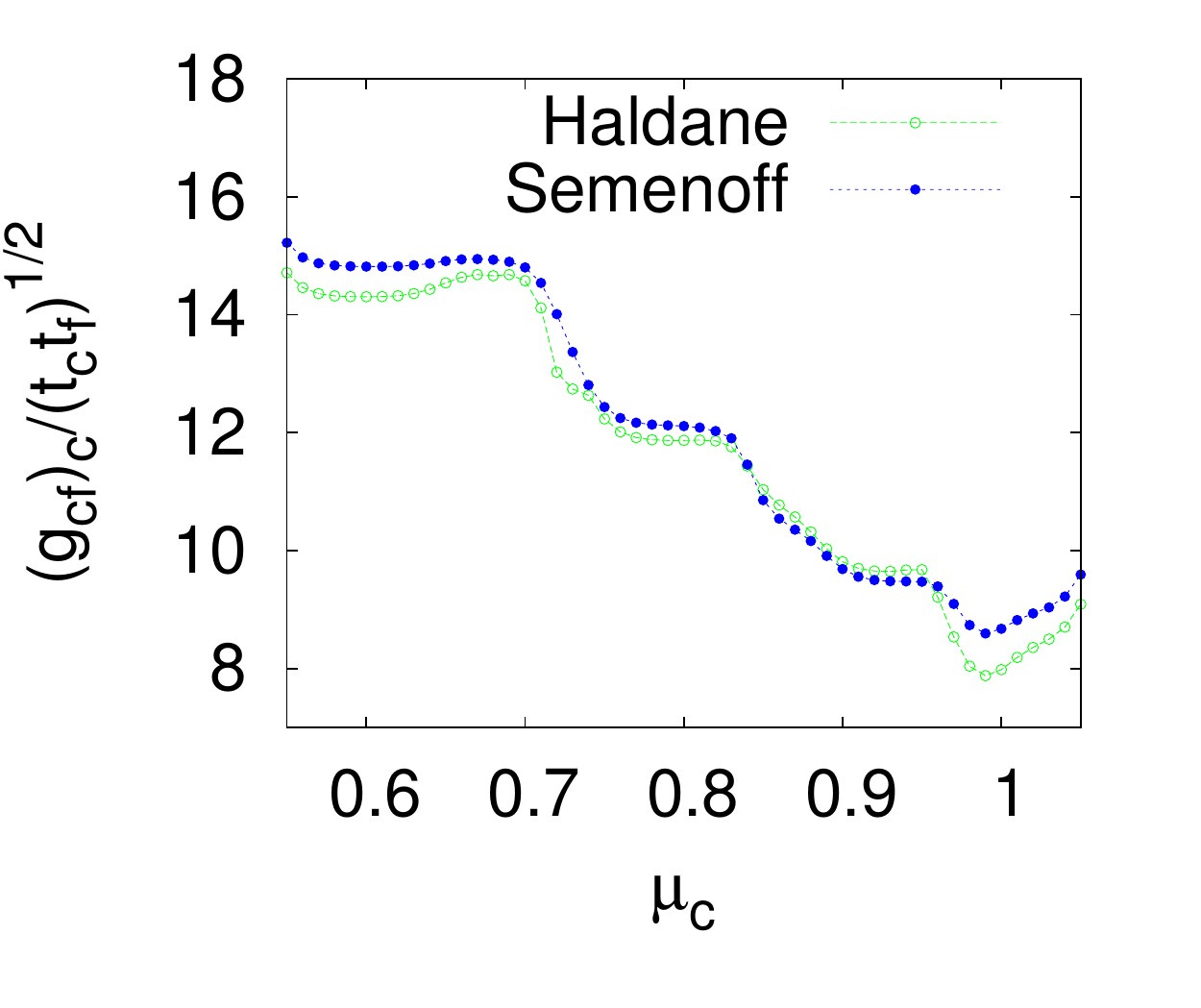}
\includegraphics[width=0.49\linewidth]{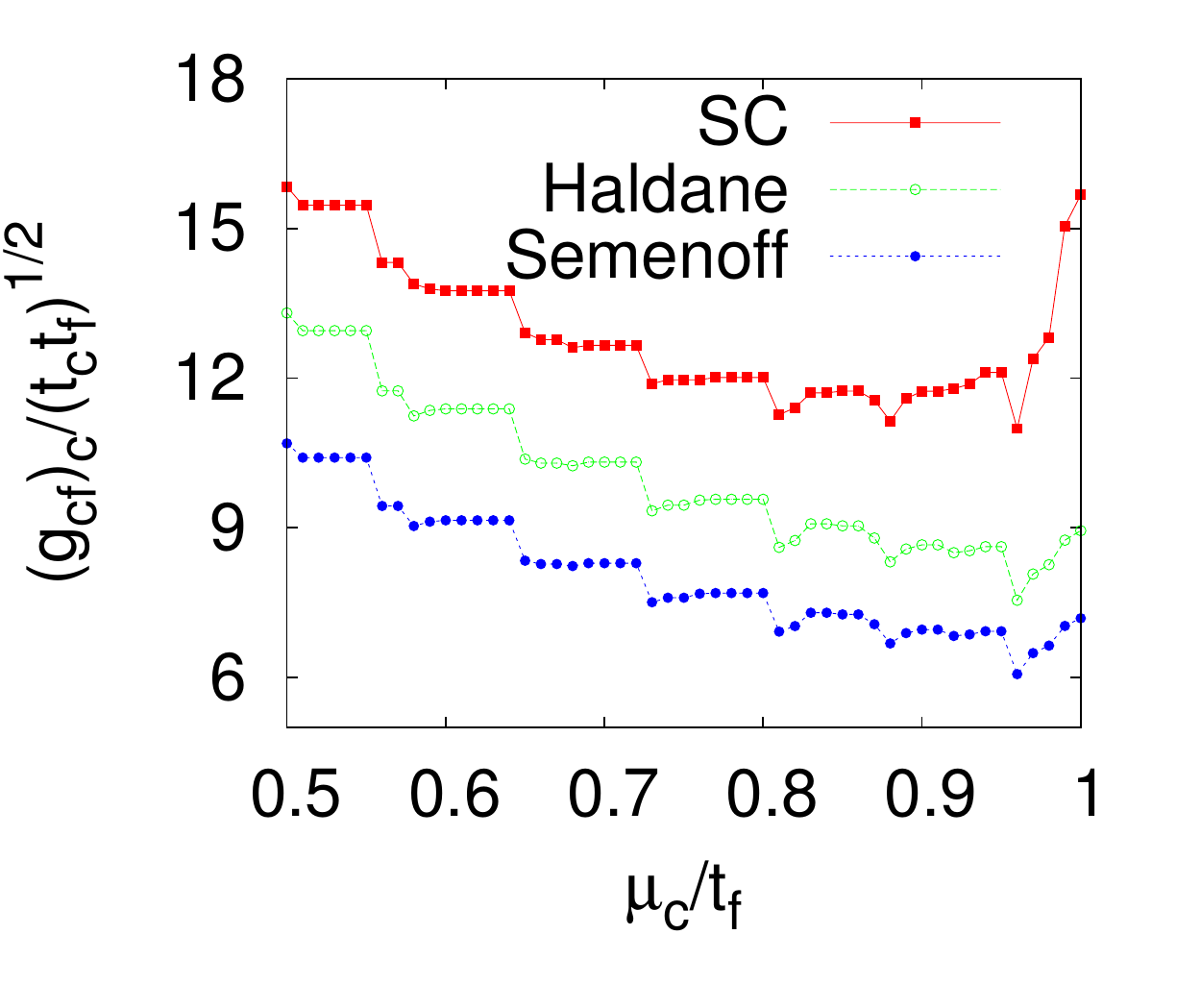}
\caption{(Color online)  Left panel: Critical value of $g_{cf}$  as a function of $\mu_c$ such that Eq.~(\ref{linear}) has a non trivial solution respectively in the odd sector (Haldane mass) and the even sector (Semenoff mass).  Right panel: Critical value of $g_{cf}$  as a function of $\mu_c$ respectively for the instability of superconductivity, Haldane phase and Semenoff phase by limiting the RKKY interaction within nearest-neighbour and next-nearest neighbour interaction. The dominance of the Semenoff phase demonstrates the necessity of the long tail of the RKKY interaction for the emergence of the Haldane phase.  }\label{gcf}
\end{figure}

At a mean-field level, we can solve the Green's function $G_f(\omega, \mathbf{p})$ by using the ansatz: $G_f(\omega, \mathbf{p})^{-1}=\omega-a(\mathbf{p})\tau_x-b(\mathbf{p})\tau_y-c(\mathbf{p})\tau_z$ such that  $a(\mathbf{p}), b(\mathbf{p}), c(\mathbf{p})\in \mathbf{R}$. Then, we are able to find self-consistent equations for the real function $a(\mathbf{p}), b(\mathbf{p}), c(\mathbf{p})$. If we denote $E(\mathbf{k})=t_f|g(\mathbf{k})|$, the projector to the lower band then becomes:
\begin{equation}\label{projector}
Pf_{JI}(\mathbf{k}+\mathbf{q})=\frac{1}{2}[1+\frac{a(\mathbf{k}+\mathbf{q})\tau_x+b(\mathbf{k}+\mathbf{q})\tau_y+c(\mathbf{k}+\mathbf{q})\tau_z }{E(\mathbf{k}+\mathbf{q})}]
\end{equation}
By inserting Eq.~(\ref{projector}) into Eq.~(\ref{action}),  we obtain three coupled non-linear equations, which are hard to resolve. By replacing $a(\mathbf{p})$ and $b(\mathbf{p})$ on the right hand side of the self-consistent equation by $\text{Re} g(\mathbf{p})$ and $\text{Im} g(\mathbf{p})$, we have the first order corrections of the bare Green's function:
\begin{widetext}
\begin{eqnarray}\label{self_consistent}
a(\mathbf{p})&=&t_f\text{Re} g(\mathbf{p})+\frac{g_{cf}^2}{2}\sum_{\mathbf{q}}\frac{\text{Re}(t_f\chi_{AB}(0, \mathbf{q}, \mu_c) g(\mathbf{p}+\mathbf{q}))}{E(\mathbf{p}+\mathbf{q})}\\
b(\mathbf{p})&=&t_f\text{Im} g(\mathbf{p})+\frac{g_{cf}^2}{2}\sum_{\mathbf{q}}\frac{\text{Im}(t_f\chi_{AB}(0, \mathbf{q}, \mu_c)g(\mathbf{p}+\mathbf{q}))}{E(\mathbf{p}+\mathbf{q})}\\
c(\mathbf{p})&=&-\frac{g_{cf}^2}{2}\sum_{\mathbf{q}}\frac{\chi_{II}(0, \mathbf{q}, \mu_c)}{E(\mathbf{p}+\mathbf{q})}c(\mathbf{p}+\mathbf{q}).
\end{eqnarray}
\end{widetext}
In Eqs.~(7) and (8), the RKKY interaction renormalizes the graphene band structure, and we have checked that the modification is of one order smaller than the function $a(\mathbf{p})$ and $b(\mathbf{p})$. In Eq.~(9), the RKKY interaction opens a gap in the system with the function $c(\mathbf{p})$. We remark that Eq.~(9) always has a trivial solution $c(\mathbf{p})=0, \forall \mathbf{p}$ for any values of $g_{cf}$. However, when $g_{cf}$ is larger than one critical value $(g_{cf})_c$ to be determined, there can exist a non-trivial solution for the function $c(\mathbf{p})$. It is reasonable to make the approximation of $E(\mathbf{k}+\mathbf{q})=t_f|g(\mathbf{k}+\mathbf{q})|$ before this instability onset. In order to study the instability onset of the function $c(\mathbf{p})$, we take the limit $c(\mathbf{p})\rightarrow0$, and therefore obtain a solvable linear equation:
\begin{equation}\label{linear}
c(\mathbf{p})=-\frac{g_{cf}^2}{2}\sum_{\mathbf{q}}\frac{\chi_{II}(0, \mathbf{q}, \mu_c)}{t_f|g(\mathbf{p}+\mathbf{q})|}c(\mathbf{p}+\mathbf{q}),
\end{equation}

We discretize the first Brillouin zone and Equation~\ref{linear} involves a real matrix with the dimension as the number of discretization points. Equation~\ref{linear} then turns into:
\begin{equation}
[V_{\chi}]=M_{\chi}[V_{\chi}],
\end{equation}
in which $V_{\chi}$ is the column vector representing the discretized  function $c(\mathbf{p})$.  The criticality occurs when the largest eigenvalue of the matrix $M_{\chi}$ attains $1$.  Eigenvectors of the matrix $M_{\chi}$ have two subspaces, the odd parity subspace $c(\mathbf{p})=-c(-\mathbf{p})$ and the even parity subspace $c(\mathbf{p})=c(-\mathbf{p})$-knowing that $M_{\chi}$ is invariant under the parity symmetry since $\chi_{II}(0, \mathbf{q}, \mu_c)=\chi_{II}(0, -\mathbf{q}, \mu_c)$. In contrast to the $t-V_1-V_2$ model \cite{t_V1_V2}, we can tune the chemical potential of the fermion $c$ in order to adjust the threshold of criticality of the emergence of the Haldane and the Semenoff mass. In Fig. \ref{gcf} left panel, we see that when $\mu_c=0.992t_c$, the critical threshold of $g_{cf}$ for the emergence of the Haldane mass reaches its minimum: $(g_cf)_c=7.88\sqrt{t_ct_f}$. 

If we denote the renormalized eigenvector for the biggest eigenvalue of the matrix $M_{\chi}$ in the odd parity sector as $VO_{\chi}$, then beyond the instability threshold $g_{cf}>(g_{cf})_c$ the Haldane mass should have the similar behavior as $VO_{\chi}$: $c(\mathbf{p})=\lambda VO_{\chi}(\mathbf{p})$. The amplitude of the Haldane mass $\lambda$ is determined by minimizing the following energy as a function of $\lambda$:
\begin{eqnarray}\label{variational}
E_0(\lambda)=&&-\sum_{\mathbf{p}}\sqrt{[a(\mathbf{p})]^2+[b(\mathbf{p})]^2+[\lambda VO_{\chi}(\mathbf{p})]^2}\\
&&+g_{cf}^2\sum_{\mathbf{p}, \mathbf{q}}\chi_{IJ}(0, \mathbf{q}, \mu_c)\frac{\lambda^2VO_{\chi}(\mathbf{p}) VO_{\chi}(\mathbf{p}+\mathbf{q})}{[E(\mathbf{p})E(\mathbf{p}+\mathbf{q})]},\nonumber
\end{eqnarray}
in which $E_0(\lambda)$ is the energy of the half-filled fermion $f$ under the RKKY interaction. The amplitude $\lambda$ is plotted as a function of $g_{cf}$ in Fig.~\ref{rkky_09} right panel.

\emph{Study of Long-Range Interaction}- Now, we study the importance of the long-range aspect in the RKKY interaction. We compare the results of Fig.~\ref{gcf} left panel with those from an effective $V_1<0, V_2>0$ model, where for each $\mu_c$, the parameters are extracted from the RKKY coupling, and we study the possible emergent instabilities. The superconducting instability entailed by the attractive nearest-neighbor interaction and the Haldane and Semenoff instability entailed by the repulsive next-nearest-neighbour interaction. The Semenoff mass, which shifts the chemical potential difference between the two sublattices, is connected to various charge density orders.  After resolution of the self-consistent relations, we see from Fig.~\ref{gcf} right panel that superconductivity is not favored among the three and the QAH (Haldane) phase is never stable. The absence of superconductivity can be justified by the linear spectrum of the Dirac fermions at half-filling, suppressing
the density of states at low energy. One needs the long tail of the RKKY interaction to frustrate the charge density wave (the Semenoff sector) from the $V_2$ interaction in agreement with recent numerical results \cite{t_V1_V2}. In this sense, the situation close to the Van Hove filling for the fast particles seems slightly better: all the longer range interaction channels are of the same order in magnitude.

\emph{Quantum Spin Hall Effect}-  Next, we also consider the case of spin-1/2 $f$-fermions with $\mu_c=0.992t_c$ and include the Hubbard interaction
\begin{eqnarray}
\begin{split}
H_f=&\sum_{\mathbf{p}, \sigma}\{[a(\mathbf{p})+ib(\mathbf{p})]f^{\dagger}_{a\mathbf{p}\sigma}f_{b\mathbf{p}\sigma}+[a(\mathbf{p})-ib(\mathbf{p})]f^{\dagger}_{b\mathbf{p}\sigma}f_{a\mathbf{p}\sigma}\}\\
&+\sum_{\mathbf{p}, I, \sigma}c_I(\mathbf{p})(f^{\dagger}_{a\mathbf{p}\sigma}f_{a\mathbf{p}\sigma'}-f^{\dagger}_{b\mathbf{p}\sigma}f_{b\mathbf{p}\sigma'})\sigma_{\sigma\sigma'}^I+H_I\\
H_I=&U\sum_if_{i\up}^{\dagger}f_{i\up}f_{i\dw}^{\dagger}f_{i\dw}.
\end{split}
\end{eqnarray}
in which $\sigma^I$ ($I=0, x, y, z$) are the Pauli matrices acting in spin space and $\sigma^0$ is the identity matrix. The functions $a(\mathbf{p})$ and $b(\mathbf{p})$ are renormalized amplitudes which are insensitive to the spin degrees of freedom. Here, the (bare) Hubbard interaction $U$ also includes a renormalization from the RKKY contribution. The topological phase here mediated by the spin degrees of freedom can also be a quantum spin Hall (QSH) phase of Kane-Mele type \cite{kane-mele}. Then, we introduce two order parameters: $c_0(\mathbf{p})$ and $c_I(\mathbf{p})$ related to the QAH and QSH phases respectively. Again, we adjust to zero the renormalized chemical potential of the $f$-fermions. Physically, if $c_0({\mathbf{q}}) \neq 0$ we are in a QAH phase, whereas when $c_I(\mathbf{q})\neq 0$ with $I=(x,y,z)$ then we are in a QSH phase. Through a careful analysis of the quantum fluctuations \cite{Raghu}, one establishes that the QSH phase is always favoured compared to the QAH phase for spinful fermions, due to the presence of Goldstone modes appearing from the breaking of the continuous rotational symmetry in the QSH phase. Therefore, we only take into account the order parameter $c_{I}(\mathbf{p})$. We find similar equations as in Eq. 9 and found that for $U=0$, the critical value $(g_{cf})_c=7.88\sqrt{t_ct_f}$ as the spinless case. By applying the slave-rotor technique \cite{FlorensGeorges}, we find that the Mott transition out of the QSH phase occurs at a relatively large $U_c$ in analogy to the Kane-Mele Hubbard model \cite{RachelLeHur}.


To summarize, we have introduced a fermion-fermion mixture in graphene-type lattices, with one fast component characterized by a large tunneling strength. We have shown that the interaction produced on the alternative species allows one to implement in realistic conditions a quantum anomalous Hall phase or a quantum spin Hall phase when we adjust the chemical potential for the fast fermion $c$ either close to $\mu_c=0.6t_c$ or close to the Van Hove filling factor, where more competing channels allow one to frustrate charge density wave orders. In particular, the long-range aspect of the RKKY interaction allows one to frustrate charge density wave orders. It should be noted that the induced relatively weak nearest-neighbor interaction here is attractive and other charge density wave orders are frustrated by the difference of chemical potentials between the fast and slow fermions, which should ensure the stability of the topological phase beyond the Hartree-Fock argument. This gives one the opportunity to observe topological Mott insulators in ultracold mixtures.

We acknowledge discussions with Doron Bergman, Maria Daghofer, Marcel Franz, Walter Hofstetter, Peter P. Orth, Tami Pereg-Barnea, Alexandru Petrescu, Stephan Rachel, Leticia Tarruell, and Wei Wu. K. L. H. has benefited from discussions at KITP Santa-Barbara, and was supported in part by the National Science Foundation under Grant No. PHY11-25915.

\pagebreak

\end{document}